\documentclass{aa}

\begin{document}
\title{On the orbital period of the cataclysmic variable V\,592 Herculis}

\author{R.E.\ Mennickent \inst{1}\fnmsep\thanks{Based on observations
obtained at the European Southern Observatory, ESO proposal 61.D-0396.}
\and  C.\ Tappert \inst{1} \and R.\ Gallardo \inst{1}
\and  H.W. Duerbeck \inst{2}
\and T. Augusteijn  \inst{3}}
\institute{Universidad de Concepci{\'o}n, Departamento de F\'{\i}sica,
 Casilla 160--C, Concepci{\'o}n, Chile\\
\email{rmennick@stars.cfm.udec.cl}
\and
University of Brussels
(VUB), Pleinlaan 2, 1050 Brussels, Belgium
\and  Isaac Newton Group
of Telescopes, Apartado 321, 38700 Cruz de La Palma,
Canary Islands, Spain}

   \offprints{R.E.\ Mennickent}
   \mail{rmennick@stars.cfm.udec.cl}
   \date{}

\abstract{We present a spectroscopic study of the long-recurrence-time
dwarf nova V\,592 Herculis based on observations obtained during
its August 1998 superoutburst.  From the analysis of  the radial
velocities of the H$\alpha$ emission line we find a most likely
orbital period of 85.5 $\pm$ 0.4 minutes, but the 91.2 $\pm$ 0.6 minutes
alias cannot be completely discarded. Both periods imply a
very small period excess
and supports the brown-dwarf like nature of the secondary star.
}

\maketitle

\section{Introduction}

Cataclysmic variable stars (CVs) are interacting binaries
consisting of a white dwarf accreting matter from a red dwarf
donor. In non-magnetic CVs, the transferred gas spirals onto
the white dwarf, forming an accretion disk. Due to the partial hydrogen ionization, the
disk  is  thermally unstable, jumping in temperature when a certain
critical density is reached. This hot state is accompanied by
increased viscosity and a release of  luminous energy when
the material rapidly drops onto the white dwarf. This event is
called an outburst. In some CVs, quasi-periodic humps are seen
in the light curve during extended outbursts; they are called
superhumps and the outbursts are called superoutbursts. According to
current theories, these peculiar CVs, the
so-called SU UMa stars, should contain low mass secondaries;
some of them  could have been eroded after a
long time of mass transfer. It is even possible that many of them
harbour secondary stars with masses less than the minimum mass
needed to sustain hydrogen fusion in their cores (e.g. Howell et al.\,2001).
Such objects should be found among ultra-short
orbital period systems ($P_{o}$ $\sim$ 80 minutes)
with very low mass transfer rates ($\dot{M} \sim 10^{15}$ g/s),
and should be characterized by rare and large-amplitude outbursts.
This subgroup of the SU UMa stars is usually called the WZ Sge stars.
Until now, a relatively small number of these objects
have been studied in detail
(Kato et al.\, 2001),
in part due to the fact that their low luminosities
make them hard to study in quiescence,
even with large aperture telescopes.
In this paper we make a contribution to the understanding of
these rare objects presenting the first spectroscopic study of
V\,592 Herculis during superoutburst.

\section{Observations and reduction}

The third historical outburst of V\,592 Her was
detected at visual magnitude 12.0 by the Finnish observer
Timo Kinnunen on 1998 August 26.835 UT.
The outburst was confirmed on August 27.181 UT by Lance Shaw in
California (see also Waagen 1998).
Observations reported to VSNET (http://www.kusastro.kyoto-u.ac.jp/vsnet/)
indicate that
the maximum occurred near the day of the detection. We observed
V\,592 Her both photometrically and spectroscopically. The
photometry,  obtained 10 days later than the spectroscopy, was
already given in Duerbeck \& Mennickent (1998), who confirmed the
presence of superhumps, and thus the SU UMa star
nature of  V\,592 Her,
and constrained the
superhump period to $SHP_{1}$ =  0.06391 $\pm$ 0.00020 days
or $SHP_{2}$ = 0.06007 $\pm$ 0.00020
days.

We obtained 40 optical spectra of
V\,592 Herculis with the EMMI spectrograph mounted at
the ESO 3.5-m NTT at La Silla Observatory,
during 1998 August 29 and 30.
3 spectra were obtained with grating \#13, providing a wavelength
range of 4010$-$9420 \AA, and 37 spectra
were obtained with grating \#8, yielding a wavelength range of
4475$-$7040 \AA. We used  slit widths of
5 and 1 arcseconds, which
yielded spectral resolutions of 5.5 and 2.5
\AA, respectively.
The standard star LTT\,7379 was observed with a 5-arcsecond
slit to flux-calibrate the spectra.
He-Ar lamp exposures were taken after
typically each hour of science exposures.
We reduced the images using standard IRAF procedures\footnote{IRAF is distributed by the National Optical
Astronomy Observatories, which are operated by the Association of
Universities for Research in Astronomy, Inc., under cooperative
agreement with the National Science Foundation.},
correcting for bias and flat field. In order to
wavelength-calibrate the spectra, we used
calibration functions obtained by fitting
$\approx$ 30 He-Ar lines with a typical standard deviation of 0.4 \AA~
(18 km/s at H$\alpha$).   An observing log is given in Table 1.
The superoutburst light curve, as constructed from the VSNET data
archive, is shown in Fig.\,1, indicating  that our spectroscopic
observations were obtained only a few days after the maximum.


\begin{table}
\caption[]{Journal of  observations. Grating,
resolution, exposure time and number of spectra are given.
The zero point for Heliocentric
Julian Day is HJD$_{0}$ = 2\,451\,000.}  \begin{center}
\begin{tabular}{ccccc} \hline
\multicolumn{1}{c}{Grating} &
\multicolumn{1}{c}{Res. (\AA)} &
\multicolumn{1}{c}{exptime (s)} &
\multicolumn{1}{c}{HJD}&
\multicolumn{1}{c}{N} \\
\hline
Gr13  &5.5 &300 &54.4826-4942  & 3\\
Gr8   &2.5 &300 &54.4996-5888  &18 \\
Gr8   &2.5 &300 &55.4884-5851  &19 \\
\hline \end{tabular}
\end{center} \end{table}

\section{Results}

\subsection{Spectrum description and variability}

The averaged spectrum for grating 13 is
shown in the upper panel of Fig.\,2.
It reveals a steep blue continuum
with hydrogen and helium absorption lines.
Spectrophotometric magnitude determination of this spectrum
yields $V$ = 14.4 and $V-R$ = 0.8.
The continuum can be approximated by a function
F$_{\lambda} \propto \lambda^{-\alpha}$,
with $\alpha$ = 3.50 $\pm$ 0.01.
This spectral energy distribution
is bluer than the  expected for an infinitely large steady state
disc, i.e. F$_{\lambda}$ $\propto$ $\lambda^{-7/3}$ (Lynden-Bell 1969).
It also differs from the observed UV spectra
of most dwarf novae during outburst, viz.\, $\approx -2$
(Verbunt 1987).
A careful examination
of the spectrum reveals the presence of emission
cores inside the H$\alpha$ and He\,I absorption lines (Fig.\,2, two lower panels).

We measured equivalent widths between two points of the continuum
separated by $\pm$ 2000 (1250) km/s from the H$\alpha$ (He\,I 5875)
rest wavelength. In the case of H$\beta$, in order to avoid
the blending with He\,I 4920, we made the measurements between continuum
points located at $-$3500 km/s and +2500 km/s from the rest
wavelength. The averaged equivalent widths and their standard deviations
were   1.6 $\pm$ 0.3, 5.2 $\pm$ 0.5 and 0.91 $\pm$ 0.15 \AA,
for H$\alpha$, H$\beta$ and He\,5875 and they did not change during the two nights.
The H$\delta$ and H$\gamma$ absorption lines seen in the grating 13
spectra had equivalent widths of 8 and 7 \AA, respectively. They
were measured as in H$\beta$ but with the upper limit wavelength of
3500 km/s. In the average grating 8 spectrum we also detected weak lines of
He\,I 4474, He\,II 4686, He\,I 4920, He\,I 5016/5048 and  He\,I 6678
(Fig.\,2 middle and bottom panel).
The He\,I 5016/5048 blend shows a central emission at 5027 \AA, while
He\,I 4920 and He\,I 5875 show emission cores at
$\lambda$ 4922 and 5874 \AA. He\,II 4686 also shows
an emission core at $\lambda$ 4687 \AA.
The strength of the emission component,
relative to their absorption component, is much stronger in
the He\,II line than
in the H and He\,I lines.
The FWHM for the double peak H$\alpha$
emission core seen in the averaged grating 8 spectrum
is 580 km/s and the peak separation 270 km/s.


\subsection{Radial velocities and the orbital period}

We measured radial velocities for several features in the spectra.
We applied the cross correlation technique to the absorption wings
and emission lines separately yielding  very noise velocity curves.
After several trials, we realized that the least noisy
radial velocities were obtained by measuring the position of the
H$\alpha$ emission core interactively with the cursor in the {\it splot}
IRAF routine. For that, we used the {\it k} key and also choose
by eye the position  where the intensity was maximum. Both methods
yielded similar results.
We searched for periodic variations in these velocities
using both the Scargle and the AOV algorithm
(Scargle 1982; Schwarzenberg-Czerny 1989) implemented
in MIDAS. The results in Fig.\,3 show possible periods at
$P_{1}$ = 0.0633 $\pm$ 0.0004 days (91.2 min), $P_{2}$ =
0.0594 $\pm$ 0.0003 days (85.5 min) and
$P_{3}$ = 0.0561 $\pm$ 0.0004 days (80.8 min). The errors correspond to
the $HWHM$ of the peaks.
In order to discriminate between the
possible frequencies and to derive the true period, we applied
the method described
in Mennickent \& Tappert (2001). In this method, sets of
radial velocity curves are generated with the same noise characteristics
and time distribution as the original dataset. These datasets were
analyzed with the idea that, after many simulations,
the true period emerges like the most recurrent period in the trial
periodograms. We fitted the data with a sine
function corresponding to the peak frequency (this was done
for the three candidate frequencies).
Then a Monte Carlo simulation was applied in such a way that a
random value from an interval consisting of $\pm$ 3 times
the sigma of the sine fit was added to each data point on the fit function.
The Scargle algorithm was applied to the resulting data set and
the highest peak was registered. The histogram for these values
is shown in Fig.\,4. The peak at 0.0594 days shows the narrowest
and highest peak, while the other periods have a broader
distribution. The total number of maxima is 327, 338 and 335 for
periods $P_{1}$, $P_{2}$ and $P_{3}$ respectively.
From the above we conclude that $P_{2}$ is slightly, but not
conclusively favoured. However, there is another line of evidence
favouring $P_{2}$. Duerbeck \& Mennickent (1998) gave two
possible orbital periods, based on the modern calibration
of the Schoembs \& Stolz relation, namely,
$P_{1}$($SHP_{1}$) = 0.06239 $\pm$ 0.00020
and $P_{2}$($SHP_{2}$) = 0.05898 $\pm$
0.00020 days. This seems to exclude $P_{3}$. In addition,
the differences between observed and predicted periods are
$P_{1}$ $-$ $P_{1}$($SHP_{1}$) = 0.0009 $\pm$ 0.0004 days
and $P_{2}$ $-$ $P_{2}$($SHP_{2}$) = 0.0004 $\pm$ 0.0004 days.
According to Mennickent et al.\, (1999), these differences
rarely exceed 0.00075 days, so $P_{2}$ is also
favoured in this case. The near coincidence between the
superhump and orbital period arises the question if the
radial velocity is being modulated by rotation
around the center of mass of the binary or, alternatively,  by the
superhump period. To our knowledge, there is no evidence for
``superhump-modulated" radial velocities in previously published
work, so here and thereafter we will assume that the
period found really reflects the binary orbital period.

The radial velocities of the H$\alpha$ emission core,
folded with the $P_{2}$ period, are shown in Fig.\,4.
The ephemeris for the red to blue passing is:\\

$T_{0} = 2\,451\,054.4969(12) + 0.0594(3)\,E $\hfill(1)\\

The figure also shows the best sine fit,
with amplitude $179 \pm 11$ km/s and zero point
$-192 \pm 15$ km/s. We obtained 161 $\pm$ 39 km/s
and 217 $\pm$ 22 km/s
for the amplitudes of first and second night,
respectively, and -165 $\pm$ 28 km/s and -195 $\pm$ 17 km/s for
the corresponding zero points, so we do not find evidence for a
$\gamma$ shift like those observed in other dwarf novae during
outburst (e.g. in VY Aqr, Augusteijn 1994 derived a shift of 155 km/s,
in TU Men Stolz \& Schoembs 1984 derived a full amplitude of $\sim$
560 km/s, in Z Cha  Hoeny et al. 1988 derived $\sim$ 160 km/s and
in TY PsA, Warner et al.\, 1989 derived $\sim$ 600 km/s).
Variations in the
system velocity have been explained with the "precessing
eccentric disc" model of Whitehurst (1998) for superhumps. In this model
$\gamma$ is expected to change with the precession period of the
disc, which is equal to the beat period between the superhump and the
orbital period. For V\,592 Her the expected precession period of the disc
is $\sim$ 5 or $\sim$ 7 days. Therefore, the non-detection
of a significant $\gamma$ shift during two consecutive nights in V\,592 Her
could be the result of our very short baseline.

The application of the ``double Gaussian"
convolution mask algorithm
(Schneider \& Young  1980, Shafter 1983) to the radial velocities
of the {\it inverted} H$\beta$ absorption
profile resulted in a large  half-amplitude radial velocity near the line center
(about 110 $\pm$ 30 km/s between 250 and 1200 km/s,
probably reflecting contamination by unseen emission) and a lower
half-amplitude of 46 $\pm$ 25 km/s in the line wings
(between 1200$-$2500 km/s from the line center, probably
indicating different gas dynamics for the emission and
absorption disc regions). The application of period searching
algorithms to these datasets yields results which seem to
exclude $P_{3}$ but are not conclusive regarding the other two
possible periods. It is generally known for dwarf novae (see, e.g. Warner 1995) that,
especially during outburst,
the radial velocity half-amplitude does not reflect the
white dwarf motion, likely due to the presence of
complex gas flow patterns in the accretion disk which are still
not well understood.
For this reason we do not intend here
to constraint the stellar masses using the binary mass
function based on the radial velocity half amplitude.




\section{Discussion}
 
According to Morales-Rueda \& Marsh (2002), the presence of
He\,II 4686 emission during high states could be
the result of disc irradiation
or emission by spiral shocks structures. Therefore,
the detection of He\,II 4686 in V\,592 Her could indicate also
the presence of spiral shocks in this short orbital period system.
Supporting this view is the fact that Baba et al. (2002)
imaged the WZ Sge disk during its 2001 superoutburst
using the He\,II 4686 line finding
arc-like structures.
Higher quality data than currently available are necessary
to confirm this suspicion in the case of V\,592 Herculis.

The period excess is defined as:                                \\

$\epsilon = \frac{P_{s}-P_{o}}{P_{o}}$\hfill(2)    \\
 
\noindent where $P_{o}$ is the orbital period of the binary and $P_{s}$
is the superhump period. The period excess is an
important observational parameter in the theory of
disk tidal instability, since it can be related to
the ratio between the stellar masses $q = M_{2}/M_{1}$:  \\

$1/\epsilon = [0.37q/(1+q)^{1/2}]^{-1}(R_{disk}/0.46a)^{-2.3}-1$\hfill(3)   \\

where $a$ is the binary separation (Patterson 2001).
Here we use  Patterson's approximation:\\

$\epsilon = 0.216(\pm0.018)q$ \hfill(4)\\

Using the orbital period $P_{2} =  85.5 \pm 0.4$ minutes, along with
the superhump period given by Duerbeck \& Mennickent
(1998),
we calculate a period excess of  $0.011 \pm 0.006$,
one of the shorter among SU UMa stars (Patterson 2001).
This period excess  implies a
mass ratio $q = 0.05 \pm 0.03$. If the white dwarf has a typical
mass of $0.7 M_\odot$, we find $M_{2} =  0.035  \pm 0.021 M_{\odot}$.
Comparing this value with  the Kumar limit for Population I stars
($\approx 0.07 M_\odot$, Kumar 1963) we find that
the secondary could be a brown dwarf like object. Only if the white dwarf
is massive ($\ga 1 M_\odot$) is the
mass ratio consistent with a
non-degenerate hydrogen-burning star.  On the other hand, if
$P_{1}$ (91.2 min) is the right period, we obtain
$\epsilon$ = 0.0096 $\pm$ 0.0070 and $q$ = 0.04 $\pm$ 0.03.
The above shows that
our result of a possible brown-dwarf like secondary in V\,592 Her
is robust against a
misidentification of the orbital period. In addition,
this finding
in agreement with the result of van Teeseling et al.\, (1999),
who arrived at this conclusion using considerations about magnitudes
at outburst and quiescence only.

\section{Conclusions}

\begin{itemize}

\item From spectroscopy obtained during superoutburst,
we have found the most likely orbital period of the dwarf nova V\,592 Her,
viz\, $85.5
\pm 0.4$ minutes. The 91.2 $\pm$ 0.5 minutes alias cannot be completely
ruled out.

\item We show that both periods, when combined with the reported
superhump period and theoretical relationships, give support for the
view that V\,592 Her harbours a brown dwarf like secondary star.

\item The detection of He\,II 4686 emission,
makes V\,592 Her  a good candidate to look for spiral shocks
during superoutburst.

\end{itemize}

\begin{acknowledgements}
We thanks the referee, Dr. Paula Szkody, for useful comments
that helped to improve a first version of this paper.
This work was supported by Grant
Fondecyt 1000324 and DI UdeC 202.011.030-1.0.
We are grateful to the VSNET
observers and the administrators
of the VSNET database for making their data available
through the WEB.
 
\end{acknowledgements}

\newpage

{\bf Figure Caption}

Fig.1: The light curve of the 1998 superoutburst.
Data are taken from the VSNET
archive. The dates of our observations are indicated by vertical dashed lines.

Fig.2: {\it Upper graph}:
flux-calibrated low resolution spectra of V\,592 Her.
{\it Middle graph}: higher resolution normalized spectrum around the
H$\beta$ line. {\it Lower graph:} same as middle graph but
around the H$\alpha$ line.

Fig.3: Scargle (top) and AOV (bottom)
periodograms of the H$\alpha$
radial velocity data. Possible frequencies
are labeled. 

Fig.4: Histogram of the maximum-peak frequency

Fig.5: The H$\alpha$ radial velocities folded with the
orbital  period. Data of the first and the second night are indicated
by squares and circles, respectively. The best sine fit is also shown.

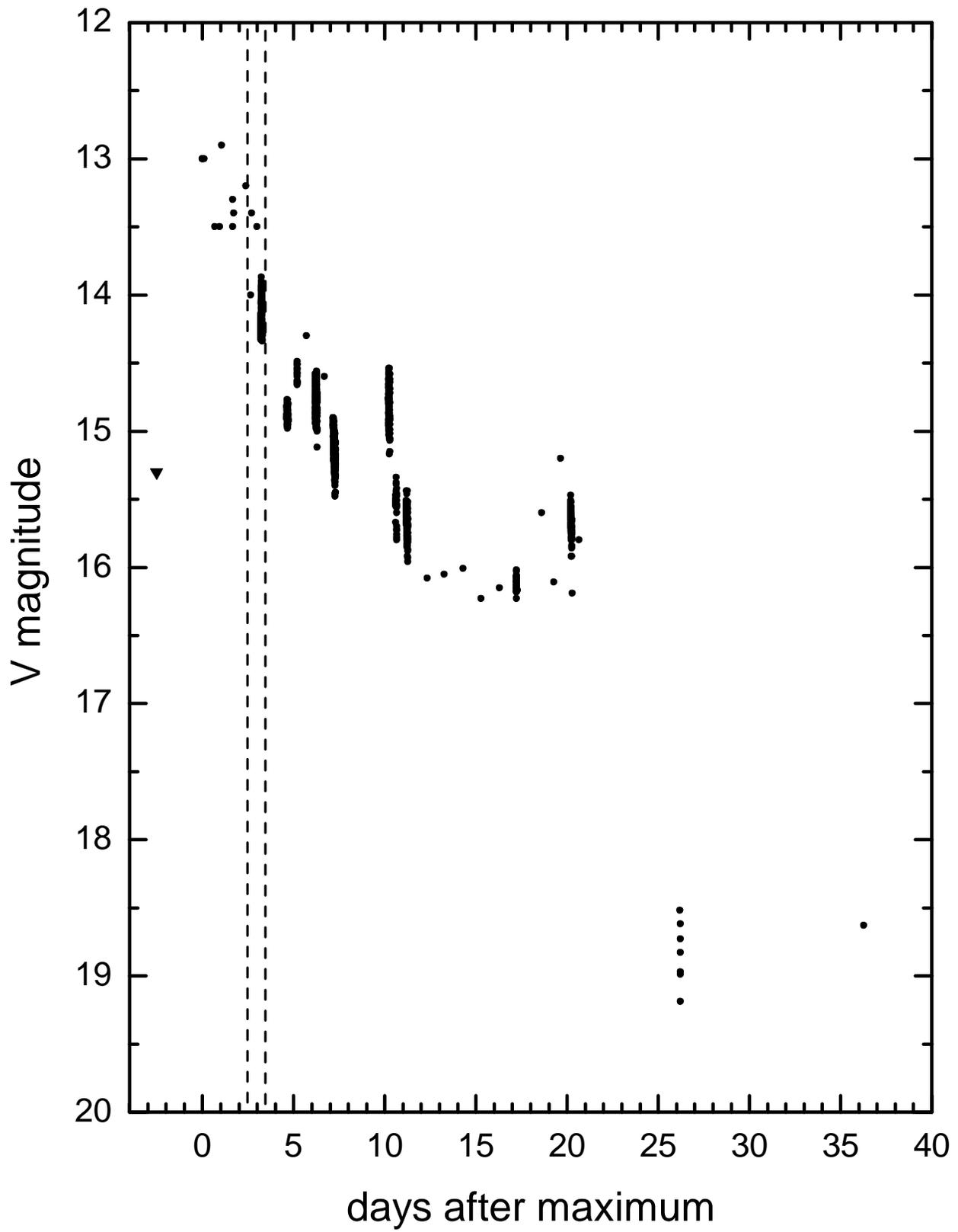
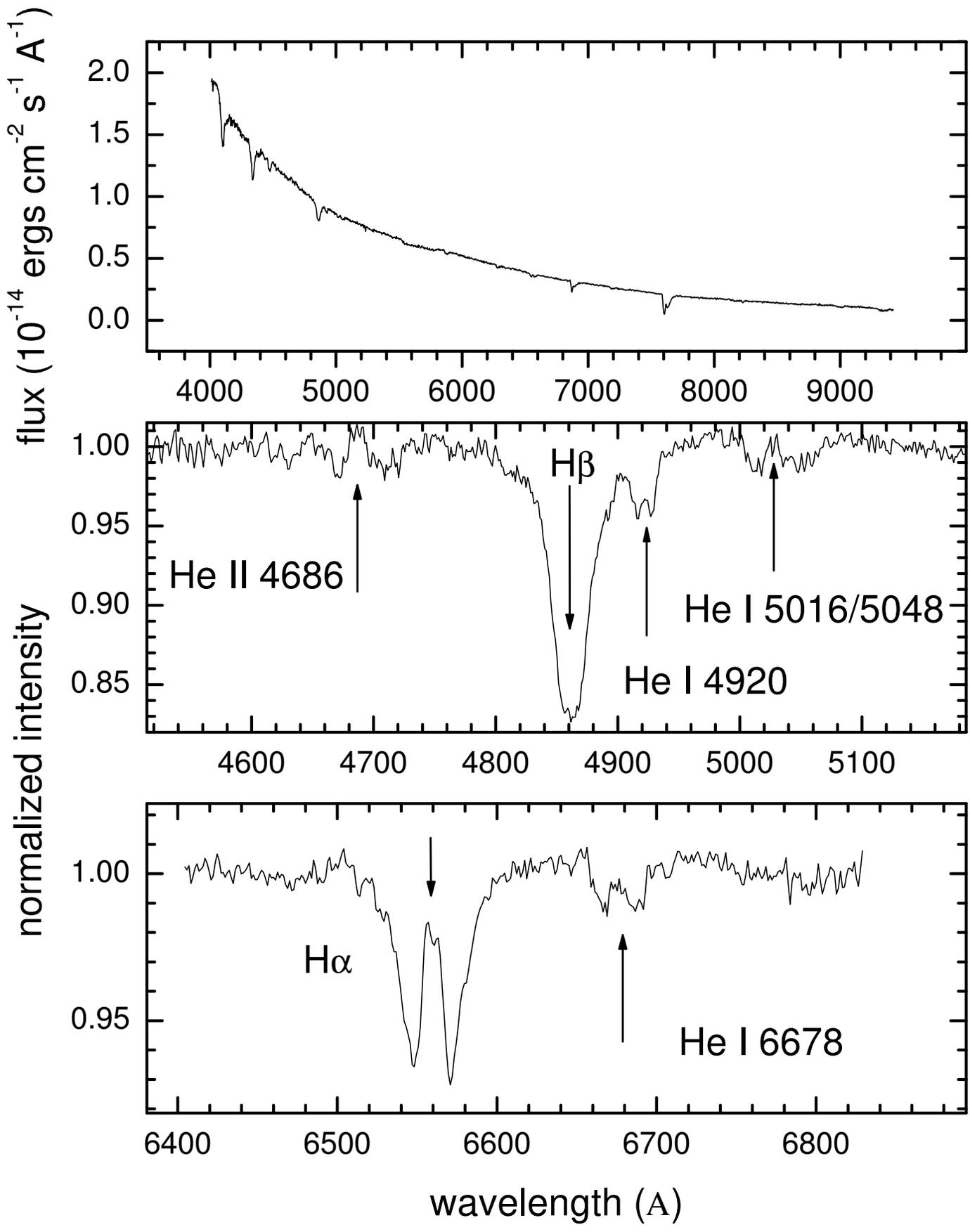
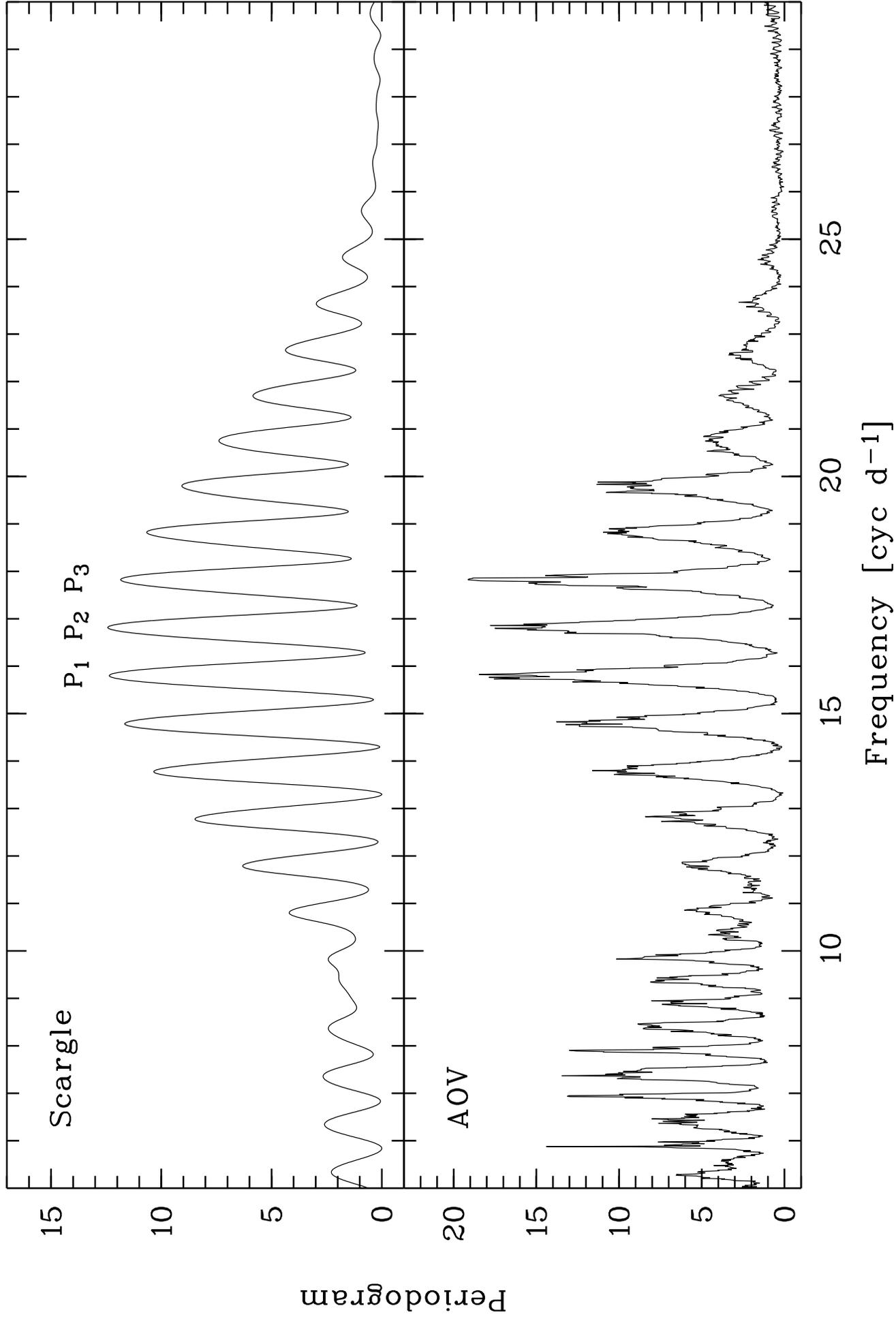
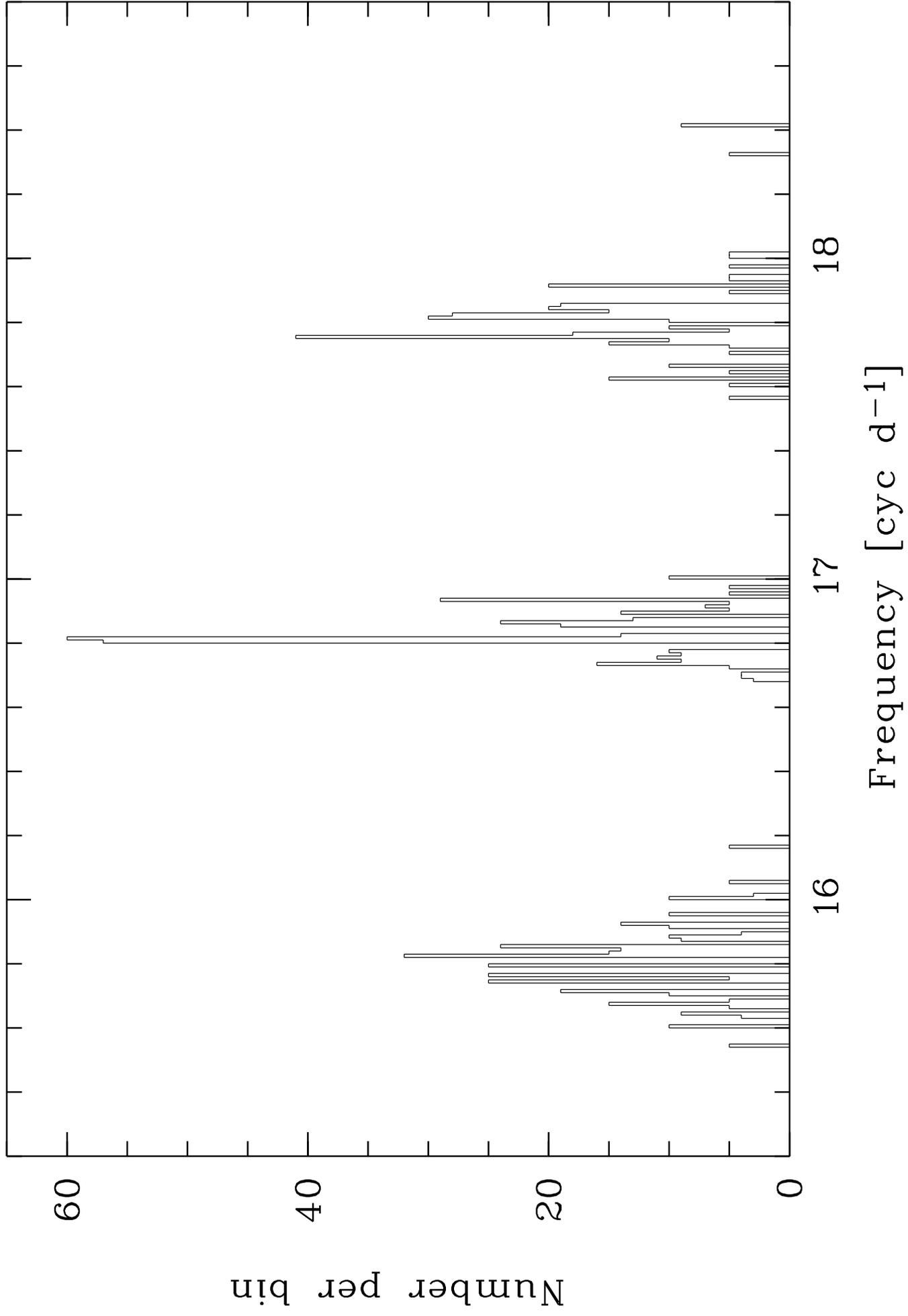
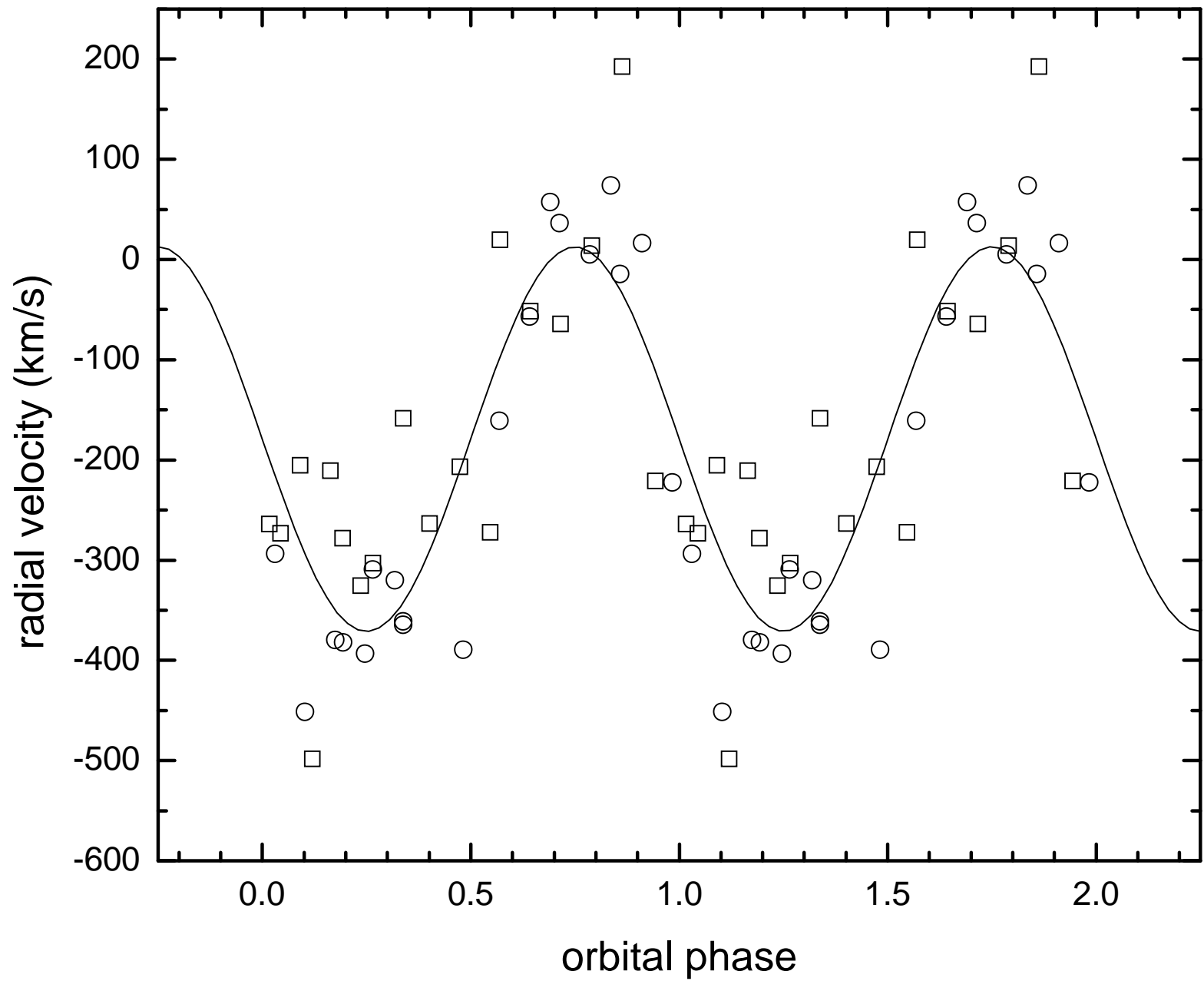
\end{document}